\documentclass[12pt]{article}
\usepackage{amssymb}
\usepackage[dvips]{graphicx}
\title{Finding Minima in Complex Landscapes:\\Annealed, Greedy and Reluctant Algorithms}
\author{ Pierluigi Contucci$^a$, Cristian Giardin\`a$^a$,\\
          Claudio Giberti$^b$ and Cecilia Vernia$^c$ \\
        \small $^a$ Dipartimento di Matematica, Universit\`{a} di Bologna\\
        \small Piazza di Porta S.Donato 5, 40127 Bologna, Italy\\
        \small {\it contucci@dm.unibo.it},
        \small {\it giardina@dm.unibo.it}\\
        \small $^b$ Dipartimento di Informatica e Comunicazione, Universit\`a dell'Insubria,\\
        \small via Mazzini 5, 21100 Varese, Italy\\
        \small {\it claudio.giberti@uninsubria.it}\\
        \small $^c$ Dipartimento di Matematica Pura ed Applicata, Universit\`{a} di Modena\\
        \small e Reggio Emilia, via Campi 213/B, 41100 Modena, Italy\\
        \small {\it vernia@unimore.it}}
\begin{document}

\maketitle
\begin{abstract}
\noindent We consider optimization problems for complex systems in
which the cost function has a multivalleyed landscape. We
introduce a new class of dynamical algorithms which, using a
suitable annealing procedure coupled with a balanced
greedy-reluctant strategy drive the systems towards the deepest
minimum of the cost function. Results are presented for the
Sherrington-Kirkpatrick model of spin-glasses.

\end{abstract}

\newpage
\section{Introduction.}
There is a standard barrier in applied science: the computational
complexity of hard (non-polynomial) problems. The modelling of
competing interactions among the components of a large system
often lead to consider the solution of a problem as the minimum of
a functional with a complex landscape. The extensive search for
the optimal configurations has a cost that grows too quickly
(usually exponentially) and become practically intractable when
the number of composing units is of the order of a few hundreds as
in the interesting cases. The study of optimizing algorithms is
then a basic step toward the solution of specific practical
problems emerging in different fields of applied science. In this
paper we build a strategy to efficiently explore the landscape of
complex functionals in combinatorial optimization in order to find
its minima both local and global. To allow the reader to better
focus on our method, let us describe the functional to be
minimized as the mathematical representation of a quickly changing
mountain profile (in large dimensions), with a high multiplicity
of local minima separated by high barriers. The a priori knowledge
of the landscape geometry is very poor and our strategy to explore
the territory in order to find good quality minima (close to the
global one) is to send signals in random directions (initial
configurations), follow their evolution according to a specified
dynamics (algorithm) and collect the observed results. Our
investigation procedure is not dissimilar from an optical
instrument in which we may tune a few parameters to better observe
the landscape and find the sites which we are interested in. The
algorithm is preliminary set by choosing the elementary dynamical
moves: this choice reflects the topology that we are associating
to our landscape and comes with a notion of vicinity and nearest
neighboring sites. The successive step is to decide the criteria
after which to select among a large multiplicity of moves. This is
done by keeping into account what we search for and what we most
fear: we want to reach the best possible minima as quickly as
possible and the worse happening is to get stuck in a local
minimum which is still far from the optimal or near optimal ones.
It appears rather intuitive that an algorithm with a too steepy
descent (greedy) has a very high risk to get stuck in poor local
minima, but at the same time a too slow descent (reluctant) would
cost a very high price in terms of computer time. It is natural to
expect, and indeed it is what we find, an optimal speed of descent
that compromise at best among having a wide exploration basin in a
reasonable amount of time. Yet the danger of remaining caught in
wrong local minima remains. To avoid it we also allow moves which
locally and momentarily deviates from the descending directions.
In other terms: to reach a good minimum it is often necessary to
overcome a high barrier. Physically, the introduction of a similar
possibility works like the availability of thermal energy where
the probability of its happening is related to the temperature of
the system: the higher the temperature the more likely are moves
upwards and viceversa. To introduce such a useful strategy we
initially allow upward and downward moves; with the time passing
the probability to go up is progressively decreased at a rate
which we may optimize (this simulates the annealing of a physical
system) and the algorithm will continue evolving according to its
downward moves. Our work and the implementation of the algorithm
is built and tested toward a standard model in combinatorial
optimization with origins in condensed matter physics: the
Sherrington-Kirkpatrick (SK) model for the mean field spin glass
phase \cite{SK,MPV}. Among the advantages of our approach, there is
the flexibility of our algorithms and their wide applicability to
practical problems like protein folding in biology \cite{MPV},
portfolio optimization in financial mathematics \cite{Bouch},
error correcting codes for digital signal transmissions
\cite{Ni}.

\section{Results.}

In the following Sections we will present details of the
Model and Algorithms we used in our simulations.
Here we summarize the main ideas and results of our analysis.

\vskip .2cm\noindent
In the Sherrington-Kirkpatrick  model the cost function is identified with
the energy of the system, the domain of the cost
function is the discrete spin configuration space
and the optimization problem amounts to find the
spin configuration with the lowest energy (ground state).
Given a proper definition of distance in the configuration
space (we can think two spin configurations to be close
if they differ only for a single spin-flip), the energy
of the system is a real-valued function forming
a complex and corrugated energy landscape,
with valleys (local minima) and peaks (local maxima).
Our optimization algorithms are described as dynamical
evolution rules in this energy landscape which,
starting from a random initial condition, drive the
system towards local minima of the energy.
The random transition from a point of the trajectory to
the successive, which is a nearest neighboring one, is
ruled by a probability with exponential density.
We consider four different algorithms: starting from the
simplest one (Algorithm 0) which allows only energy-decreasing
trajectories, we implement a sequence of refinements
(Algorithms 1,2,3) leading to more efficient strategies,
which exploit also increases in the cost function.

\vskip .2cm\noindent
With Algorithm 0 the cost-decreasing trajectory ends up as soon as it reaches
a configuration which, according to our notion of vicinity (see Sec.~3)
is a local minimum. The parameter controlling the transition probability
function tunes the steepness of descents, generating a continuum of behaviors
ranging from a reluctant-type dynamics (very small jumps and slow convergence)
to a greedy-type one (very large jumps deep into a valley).

\vskip .2cm\noindent
A first improvement of this strategy, implemented in Algorithms 1 and 2,
is obtained by introducing a ``temperature'' in the system, which
enables random positive fluctuations of the cost function.
This is obtained through the choice of a transition probability which
gives a non zero weight to upwards moves.
With these choices we have the following scenario for Algorithms 1 and 2: the
dynamics starts with a given initial temperature and equal probability of
positive and negative moves. As the time goes on, the system is gradually
cooled until it reaches a state in which positive fluctuations are forbidden
and the dynamics continues as either greedy or reluctant,
depending on the initial temperature. With a high initial temperature the
long term behavior of the dynamics will be greedy-like,
while a low initial temperature will lead to reluctant-type motion.
The difference between Algorithm 1 and 2 lies in the convergence criterium:
while the former stops when the first local minimum is attained
(likewise Algorithm 0), the latter allows the trajectory
to escape from it in view of the possibility to reach deeper minima
(supplementary stopping conditions are required in this case).

\begin{figure}[ht!]
    \setlength{\unitlength}{1cm}
          \centering
               \includegraphics[width=13cm,height=10cm]{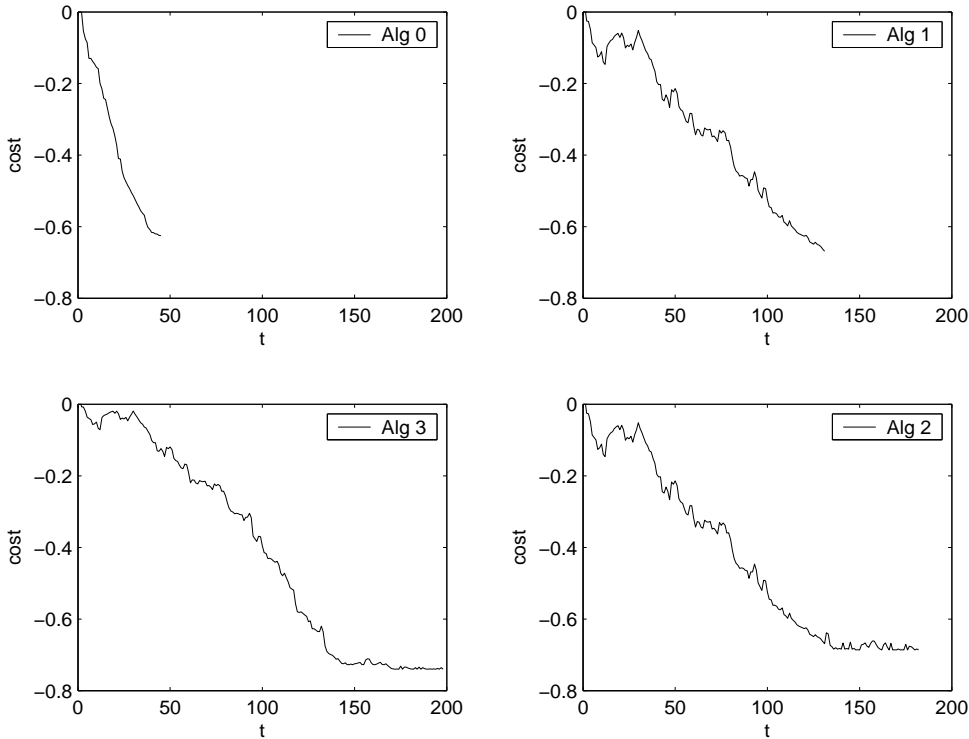}
               \caption{Typical trajectories to reach a local minimum
               configuration for Algorithm 0, 1, 2 and 3. Note that the trajectories
               generated by Algorithms 1 and 2 coincide until the first minimum is reached.}\label{fi:trai}
\end{figure}

\vskip .2cm\noindent
A further improvement of the algorithm efficiency is obtained with
Algorithm 3. In this case, the transition probability is designed to model an
initially hot system with high probability of positive moves, which is
gradually quenched; when the system is cool, positive fluctuations are
absent and the decreasing trajectories are forced to follow greedy-like
paths. In Fig.~\ref{fi:trai} typical trajectories for the four different
algorithms are reported.

\vskip .2cm\noindent
The efficiency of the algorithms are quantified on one hand by
measuring the average time needed to reach a local minimum, on the
other hand by the quality of the found minima (i.e. how deep they are).
The optimization is done by tuning the parameters which control the
transition probabilities; in particular, for Algorithms 1 and 2 this
parameter is mainly the initial temperature, while for
Algorithm 3 it is the rate of the quench, i.e. the speed of
convergence to zero of the temperature of the system.
As one would expect, for low initial temperatures
(very low possibility of energy increase),
Algorithm 1 and 2 behaves very much as Algorithm 0.
However  their differences become effective for
sufficiently high initial temperatures.
Obviously, allowing positive jumps  and escapes from local minima,
the relaxation times increase passing from Algorithm 0 to Algorithm 2;
less trivially, numerical results show that the scaling of the execution
times with respect to the system size is greatly enhanced.
This is an important fact, because it suggests that a crossover
between computation times is to be expected for systems with larger sizes.
As regards the lowest values found, similar conclusions can be drawn: going
from Algorithm 0 to Algorithm 2 deeper minima are attained.
\vskip .2cm\noindent
Algorithm 3 can be consistently compared with Algorithm 2, which is the best
performing among the first three. The computation times and their scaling with the size are similar
for the two algorithms when the initial temperature (for Algorithm 2) is high, but a clear
enhancement is obtained by Algorithm 3 when it is low. Also the minimal values of the cost functional
are similar for high temperatures, while they are lower for Algorithm 2 with low initial temperatures.
The previous remarks refer to an experimental protocol in which the search for low cost configurations
is performed testing a fixed number of trajectories. The minimization of cost at fixed elapsed computer time is
another relevant criterium for the comparison of the algorithms. In this case the best result is obtained
with Algorithm 3, even though Algorithm 2 gives comparable results.

\section{The model and the algorithms.}
\vskip .1cm
\subsection{The Sherrington Kirkpatrick model}
\vskip .5cm
\noindent
The system we  study is the Sherrington-Kirkpatrick model
of spin-glasses \cite{SK}. It is defined by the Hamiltonian
\begin{equation}
H(J,\sigma)=-\frac{1}{\sqrt{N}}\sum_{1\leq i<j \leq N} J_{ij}\sigma_i\sigma_j
\end{equation}
where $\sigma_i=\pm 1$ for $i=1,\ldots ,N$ are Ising spin variables
which interact through couplings $J_{ij}$.
These are gaussian random variables, independent and
identically distributed  with zero mean and variance $1$.
The random sign (and strength) of the interaction
generates frustration in the system, i.e. the fact
that in low energy configurations some of the couples
will have unsatisfied interaction. In particular, the ground state of the
system is far from the standard ground state of
ferromagnetic models, where all spins point in the
same direction.
The model has been solved through the replica symmetry
breaking ansatz by G. Parisi \cite{MPV}, while the rigorous solution
is still a debated issue in the mathematical physics
community. From the numerical point of view, the model
poses amazing difficulties and indeed it is often
presented as the standard example of NP-problems.
Several numerical studies have tried different
algorithms in the search of ground-state
energies, for example gradient descendent
\cite{BP,CMPP}, simulated annealing \cite{KGV,GSL},
genetic algorithms \cite{BKM,P},
extremal optimization \cite{BP01,BS,B}.
In a previous paper we developed a new
numerical scheme, which is based on a smooth
interpolation between greedy and reluctant dynamics
\cite{BCDG,BCGGUV,CGGUV1}. Here we make a further step
by proposing a new class of algorithms which we
describe in detail in the following.

\vskip .5cm
\subsection{Dynamical Algorithms}
\vskip .5cm
\noindent
We focus our attention on stochastic dynamics that
generates a sequence of spin configurations ending up on a local
energy minimum. The smooth interpolation between greedy and
reluctant dynamics studied in a previous work \cite{CGGUV1}
follows an energy-decreasing trajectory and
terminates in the first local minimum it encounters: only
transitions corresponding to a decrease in the cost (energy) function are
allowed by the algorithm. In the same spirit of Simulated
Annealing strategies \cite{KGV}, where a slow decrease of the temperature
leads the system through successive metastable states with
lower and lower energy, we think of a class of algorithms which also
accept, in some limited way, transitions corresponding to an
increase in the cost function.
In fact, these algorithms are based on the statistical
properties of metastable states: they are organized with some structure
so that the evolution dynamics can be
considered as the overlapping of a ``fast'' motion in the basin of
attraction of a local minimum and of a ``slow'' motion with jumps between
minima (the time of the dynamics is determined by the energy barriers between
these metastable states).
\vskip .2cm\noindent
In the algorithms that we are going to introduce, the transition between
the spin configuration at time $t$,
$\sigma(t)=(\sigma_1(t),\ldots,\sigma_N(t))$,
and the successive configurations
at time $t+1$, $\sigma(t+1)=(\sigma_1(t+1),\ldots,\sigma_N(t+1))$
depends on the spectrum  of energy changes of $\sigma(t)$,
obtained by flipping the spin in position $i$, for $i=1,\ldots,N$:
\begin{equation}
\Delta E_i=\sigma_i(t)\sum_{j\ne i} J_{ij}\sigma_j(t).
\end{equation}
Let also define $\Delta E_{\bar i}=\min_{1\le i\le N}\Delta E_i$ that
will be used in what follows.
\vspace{.2cm}
\noindent
As a first step, let us briefly recall the algorithm studied in
\cite{CGGUV1}, where only energy decreasing trajectory are considered.
It is described by the following procedure:

\vskip .3cm
{{\bf Algorithm 0}}
\vskip .2cm
\noindent
\begin{enumerate}
\item Initialization: choose an initial spin configuration
$\sigma(0)$ and a parameter value for $\lambda > 0$.
\item Generate a random
number $D$ with probability density
\begin{equation}
f(x)=\left\{ \begin{array}{ll}
   \lambda e^{\lambda x} & \textrm{if } x\leq 0\\
   0  & \textrm{if }  x > 0
\end{array}\right.
\end{equation}
\item Select the site $i^{\star}$ associated with the closest energy
change to the value $D$, i.e.:
\begin{equation}
i^{\star} \quad : \quad | \Delta E_{i^{\star}}-D |=\min_{i\in\{1,\ldots,N\}} \{ |\Delta E_i-D|:\Delta
E_i<0\}.
\end{equation}
\item Flip the spin on site $i^{\star}$:
\begin{equation}
\sigma_i(t+1)=\left\{ \begin{array}{ll}
  -\sigma_i(t) & \textrm{if } i=i^{\star}\\
   \sigma_i(t) & \textrm{if } i\neq i^{\star}.
\end{array}\right.
\end{equation}
\item
If $\Delta E_i > 0$, $\forall i=1,\ldots,N$, then the algorithm
stops ($\sigma(t)$ is a local minimum); otherwise repeat
from step 2.
\end{enumerate}
The dynamics generated by this algorithm follows a $1$-spin flip
decreasing energy trajectory and arrives at a configuration whose
energy cannot be decreased by a single spin-flip. The control
parameter $\lambda$ in the probability distribution function for
the move acceptance, tunes the speed of convergence to local energy
minima: the larger is $\lambda$, the bigger is the probability
of doing small energy-decreasing steps, so that the trajectory
will follow an evolution path close to level curves (reluctant)
while, small values of $\lambda$ enrich the probability of large
negative energy steps (greedy), which will quickly drive the dynamics to the
end-point.

\vskip .5cm
\noindent
As a modification of Algorithm $0$ we consider two new algorithms
(Algorithm 1 and Algorithm 2). They generate a dynamics that follows
a $1$-spin flip trajectory that, in addition to energy-decreasing transitions,
accepts also energy-increasing transitions with probability exponentially
decreasing in time.  The difference between the two is that while
the trajectory of Algorithm 1 ends up in the first local minimum
it encounters, in Algorithm 2 it may continue to explore the space
of configurations through the visit of subsequent local minima.


\vskip .3cm
{{\bf Algorithm 1}}
\vskip .2cm
\begin{enumerate}
\item Initialization: choose an initial spin configuration
$\sigma(0)$ and parameter values $0 < c_1(0) < \lambda_1$,
$0 < c_2 < \lambda_2(0)$, with the obvious constraint
\begin{equation}
\frac{c_1(0)}{\lambda_1} + \frac{c_2}{\lambda_2(0)} = 1
\end{equation}
In our simulation we chose $\lambda_1$ as the only
free parameter, by taking $\lambda_2(0) = \lambda_1$,
$c_1(0) = \lambda_1 /2$, $c_2 = \lambda_1/2$.
This amounts to  start with an equal probability of
energy decreasing and energy increasing transitions
($c_1(0)/\lambda_1 = c_2/\lambda_2(0) = 1/2$).
\item Generate a random
number $D$ with probability function
\begin{equation}
f_t(x)=\left\{ \begin{array}{ll}
   c_1(t) e^{\lambda_1 x} & \textrm{if } x\leq 0\\
   c_2    e^{-\lambda_2(t) x}  & \textrm{if }  x > 0
\end{array}\right.
\end{equation}
\item Select the site $i^{\star}$
associated with the closest energy change to the value $D$ and
with the same sign, i.e.:
\begin{equation}
i^{\star}\quad : \quad | \Delta E_{i^{\star}}-D |=\min_{i\in\{1,\ldots,N\}}
\{ |\Delta E_i-D|:\Delta E_i\cdot D >0\}.
\end{equation}
\item Flip the spin on site $i^{\star}$:
\begin{equation}
\sigma_i(t+1)=\left\{ \begin{array}{ll}
  -\sigma_i(t) & \textrm{if } i=i^{\star}\\
   \sigma_i(t) & \textrm{if } i\neq i^{\star}.
\end{array}\right.
\end{equation}
\item
If $\Delta E_i > 0$, $\forall i=1,\ldots,N$, then the algorithm
stops ($\sigma(t)$ is a local minimum).
Otherwise, change the parameter $\lambda_2(t)$ of the
probability distribution in step 2 with
a suitable  scheduling, for example
\begin{equation}
\label{eq:sch1}
\lambda_2(t)=\frac{\lambda_2(0)}{k^t}, \quad\quad\quad\quad 0<k<1
\end{equation}
and return to step 2.
\end{enumerate}
The trajectory generated by Algorithm 1 wonder in the energy
landscape (by a succession of moves which decrease and increase
energy) till it arrives to a local minimum. Starting from a symmetric
probability distribution for the spin-flip selection, as time goes on
the probability of energy-increasing moves is decreased by the
update rule (\ref{eq:sch1}).

\vskip .5cm
\noindent
Next, we want to consider an algorithm as the previous one
but with the possibility of exploring subsequent minima.
The problem one has to solve is to give an efficient criterium
to stop the dynamics. We considered the following implementation:


\vskip .3cm
{{\bf Algorithm 2}}
\vskip .2cm
\begin{enumerate}
\item Initialization: as in Algorithm 1.
Set also $m=1000$ and $\epsilon = 10^{-4}$.
\item Generate a random number $D$ as follows:

with probability function
\begin{equation}
f_t(x)=\left\{ \begin{array}{ll}
   c_1(t) e^{\lambda_1 x} & \textrm{if } x\leq 0\\
   c_2    e^{-\lambda_2(t) x}  & \textrm{if }  x > 0
\end{array}\right. \qquad \textrm{if } \quad
\frac{c_1(t)}{\lambda_1} \le m \frac{c_2}{\lambda_2(t)}\label{eq:ftAlg2}
\end{equation}
and with probability function
\begin{equation}
f(x)=\left\{ \begin{array}{ll}
   \lambda_1 e^{\lambda_1 x} & \textrm{if } x\leq 0\\
   0  & \textrm{if }  x > 0
\end{array}\right. \qquad \textrm{if } \quad
\frac{c_1(t)}{\lambda_1} > m \frac{c_2}{\lambda_2(t)}
\end{equation}
\item Select the site $i^{\star}$
associated with the closest energy change to the value $D$ and
with the same sign, i.e.:
\begin{equation}
i^{\star}\quad : \quad | \Delta E_{i^{\star}}-D |=\min_{i\in\{1,\ldots,N\}}
\{ |\Delta E_i-D|:\Delta E_i\cdot D >0\}.
\end{equation}
\item Flip the spin on site $i^{\star}$:
\begin{equation}
\sigma_i(t+1)=\left\{ \begin{array}{ll}
  -\sigma_i(t) & \textrm{if } i=i^{\star}\\
   \sigma_i(t) & \textrm{if } i\neq i^{\star}.
\end{array}\right.
\end{equation}
\item
If $\Delta E_i > 0$, $\forall i=1,\ldots,N$, and $P_t(D\ge\Delta E_{\bar i})<
 \epsilon$
then Stop.

$D$ is a random number, $P_t$ is the cumulative function of the
probability described in step 2 and $\epsilon$ is a small parameter. In other
words, if we arrive in a minimum and the probability
of a significant energy increasing transition from this local
minimum is too small (or even zero when the energy
increases are forbidden, see step 2), then the algorithm stops.
\item Change the probability distribution (\ref{eq:ftAlg2}) with
the scheduling (\ref{eq:sch1}) for $\lambda_2(t)$
(the same scheduling used in Algorithm 1) and return to step 2.
\end{enumerate}
As in Algorithm 1, the dynamics generated by this algorithm follows a $1$-spin flip
trajectory making a combination of upwards and downwards moves.
However, in this case, the trajectory
does not end up in the first $1$-spin flip
stable configuration  it encounters, at least as long
as the probability of positive moves ($c_2/\lambda_2(t)$) remains
greater than  a certain threshold ($1/m$ times the
probability of negative moves $c_1(t)/\lambda_1$ - in our experiments
$m=1000$). With this
strategy it is possible to escape from the local minima to explore the
neighboring space in view of (possible) lower energy minima. When
the probability of energy increases exceed this fixed threshold,
from this point on, only decreases in energy are accepted and so
the process terminates when the subsequent local minimum is
reached. In fact, when the process starts at time $t=0$ we choose
equal probabilities $c_1(0)/\lambda_1$ and $c_2/\lambda_2(0)$ of
cost-decreasing or cost-increasing moves, respectively, by
settling $c_2=\lambda_2(0)/2$. As the algorithm continues its
execution, we decrease $c_2/\lambda_2(t)$ towards zero, varying
the control parameter $\lambda_2(t)$ in accordance with the above mentioned
law (\ref{eq:sch1}):
$$
\lambda_2(t)=\frac{\lambda_2(0)}{k^t}, \qquad \lambda_2(0)=
\lambda_1,\quad 0<k<1
$$
(and keeping fixed $\lambda_1$) until $\frac{c_1(t)}{\lambda_1} \le
m \frac{c_2}{\lambda_2(t)}$; as a consequence, the probability of
energy-decreasing move acceptance $c_1(t)/\lambda_1$ tends to one
($c_1(t)=\lambda_1(1-c_2/\lambda_2(t))$). Therefore, while the
speed of convergence to the local energy minima is mainly tuned by
$\lambda_1$, the vanishing velocity of the probability of
energy-increasing steps is governed by the parameter $k$. Of
course, large $\lambda_1$ (and $\lambda_2(t)$) lead to evolution
paths generated by small (in absolute value) energy changes ({\it
annealed reluctant dynamics}) and the closer $k$ is to $1$, the
slower $\lambda_2(t)$ grows and then the more energy increases are
enabled. When $\frac{c_1(t)}{\lambda_1} > m
\frac{c_2}{\lambda_2(t)}$ the dynamics continues governed only by
the parameter $\lambda_1$, not depending on $t$.
\vskip .2cm\noindent
We see that for Algorithm
$2$ the possibility to escape from the minima is effective  only when $\lambda_1$ is
sufficiently small (say $\lambda_1\simeq 1$, and then $\lambda_2(0)\simeq 1$,
see (\ref{eq:sch1})). For greater values of $\lambda_1$
the possibility to explore successive
minima is not exploited and both the dynamics $1$ and $2$ can be expected to give similar results
in terms of achieved minimum energy level.
In these cases, the dynamics generated by Algorithm $2$
ends up naturally, after $t'$ steps, in the first minimum it encounters, because the
(step dependent) probability $P_{t'}$
to escape from this configuration is too small; therefore, we
expect that for large values
of $\lambda_1$ Algorithms $1$ and $2$ should be equivalent.

\vskip .5cm\noindent
Since for these algorithms the speed of convergence to the finale state is
governed by the probability function
$f_t(x)$, we can consider a third algorithm in which
the time dependence is present only in the control parameters
$\lambda_i(t)$, $i=1,2$; in this case, starting from a (in general) non symmetric
probability function, the dynamics evolves gradually
towards a final scenario in which the
system is cooled by tuning the control parameter $\lambda_1(t)$.


\vskip .3cm
{{\bf Algorithm 3}}
\vskip .2cm
\begin{enumerate}
\item Initialization: choose an initial spin configuration
$\sigma(0)$ and parameter values $\lambda_1(0)$, $\lambda_2(0)$
such that $1/\lambda_1(0) +1/\lambda_2(0) = 1$.
Set also $m=1000$ and $\epsilon = 10^{-4}$.
\item Generate a random number $D$ as follows:

with probability function
\begin{equation}
f_t(x)=\left\{ \begin{array}{ll}
   e^{\lambda_1(t) x} & \textrm{if } x\leq 0\\
   e^{-\lambda_2(t) x}  & \textrm{if }  x > 0
\end{array}\right. \qquad \textrm{if } \qquad
\frac{1}{\lambda_1(t)} \le m \frac{1}{\lambda_2(t)}\label{eq:ftAlg3}
\end{equation}
and with probability function
\begin{equation}
f(x)=\left\{ \begin{array}{ll}
   \lambda_1 e^{\lambda_1 x} & \textrm{if } x\leq 0\\
   0  & \textrm{if }  x > 0
\end{array}\right. \qquad \textrm{if } \qquad
\frac{1}{\lambda_1(t)} > m \frac{1}{\lambda_2(t)}
\end{equation}
\item Select the site $i^{\star}$
associated with the closest energy change to the value $D$ and
with the same sign, i.e.:
\begin{equation}
i^{\star}\quad : \quad | \Delta E_{i^{\star}}-D |=\min_{i\in\{1,\ldots,N\}}
\{ |\Delta E_i-D|:\Delta E_i\cdot D >0\}.
\end{equation}
\item Flip the spin on site $i^{\star}$:
\begin{equation}
\sigma_i(t+1)=\left\{ \begin{array}{ll}
  -\sigma_i(t) & \textrm{if } i=i^{\star}\\
   \sigma_i(t) & \textrm{if } i\neq i^{\star}.
\end{array}\right.
\end{equation}
\item
If $\Delta E_i > 0$, $\forall i=1,\ldots,N$, and $P_t(D\ge\Delta E_{\bar i})< \epsilon$
then Stop (as in Algorithm 2) .
\item Change the probability distribution defined in (\ref{eq:ftAlg3}) with
the same  scheduling for $\lambda_2(t)$ used in Algorithm 2
and return to Step 2.
\end{enumerate}

\begin{figure}
    \setlength{\unitlength}{1cm}
          \centering
               \includegraphics[width=13cm,height=10cm]{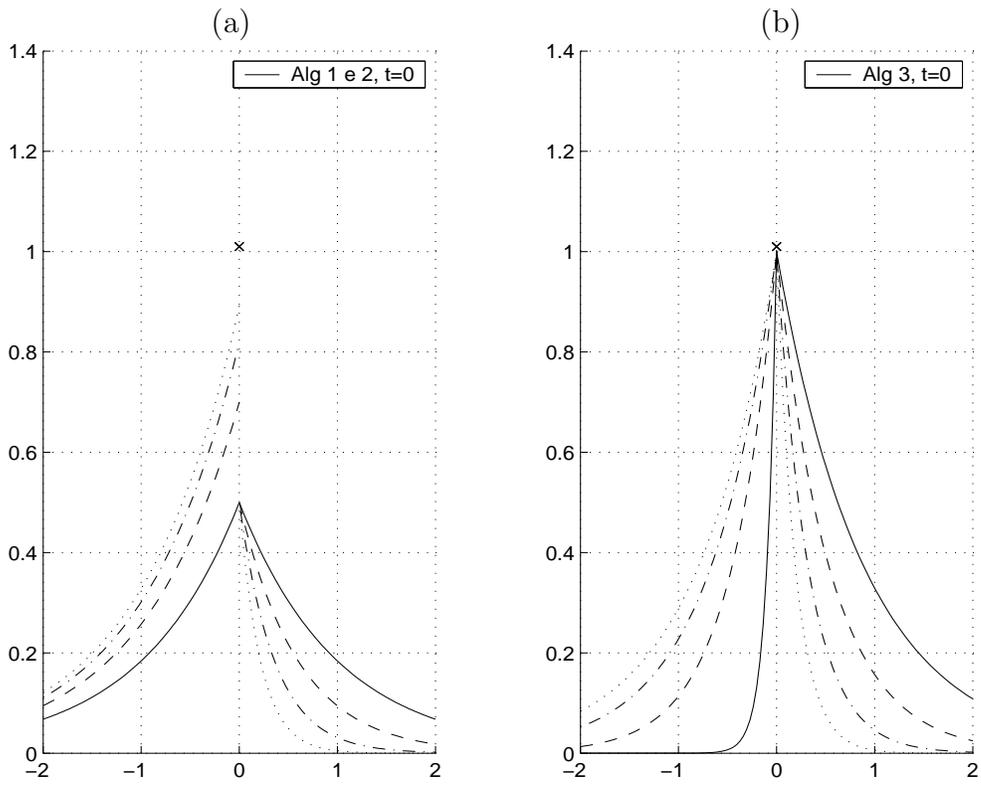}
               \put(-10.3,10){(a)}
               \put(-3,10){(b)}
               \caption{Probability density functions for Algorithm 1 and 2
               (part (a)) and for Algorithm 3 (part (b)) for different values of
               time $t$. The continuous lines refer to $t=0$; the time goes on
               passing from broken lines to dotted ones.}\label{fi:PDFdy}
\end{figure}

\noindent
The main difference between Algorithm 2 and Algorithm 3 is that
in the latter, when the process starts at time $t=0$ we have
(if $\lambda_1(0)\ne 2$) different probabilities of energy-decreasing moves
($1/\lambda_1(0)$) and of energy-increasing moves ($1/\lambda_2(0)$).
As Algorithm 3 continues its
execution, we decrease $1/\lambda_2(t)$ towards zero, varying
the control parameter $\lambda_2(t)$ in accordance with the scheduling:
\begin{equation}
\lambda_2(t)=\frac{\lambda_2(0)}{k^t}, \qquad \lambda_2(0)=
\frac{\lambda_1(0)}{\lambda_1(0)-1},\quad 0<k<1
\end{equation}
until $\frac{1}{\lambda_1(t)} \le m \frac{1}{\lambda_2(t)}$;
as a consequence, the probability of energy-decreasing move acceptance
$1/\lambda_1(t)$ tends to one
($\lambda_1(t)=\frac{\lambda_2(t)}{\lambda_2(t)-1}$). Therefore, while the
speed of convergence to the final state is mainly tuned by the initial value
$\lambda_1(0)$ of the time dependent parameter $\lambda_1(t)$ (which
tends to $1$, as time $t$ increases), the vanishing velocity of the probability of
energy-increasing steps is governed by the parameter $k$.
When $\frac{1}{\lambda_1(t^*)} > m
\frac{1}{\lambda_2(t^*)}$ the dynamics continues, for $t>t^*$, governed only by
the parameter $\lambda_1=\lambda_1(t^*)$ (close to 1) not depending on $t$.
The dynamic evolution of the probability density functions for Algorithm 1 and
2 compared with Algorithm 3 is reported in Fig.~\ref{fi:PDFdy}.

\vskip .5cm\noindent
Summarizing: the control parameters are $\lambda$ for Algorithm 0,
$\lambda_1$ and $k$ for Algorithms $1$ and $2$,
and $\lambda_1(0)$ and $k$ for Algorithm $3$.
Varying them we study the efficiency of the algorithms
by measuring the average time  to reach a metastable
configuration and the lowest energy value found
for different system sizes.

\section{Data analysis.}

To compare these annealed algorithms with those carried out in
previous works \cite{BCDG,BCGGUV,CGGUV1} and in particular with
Algorithm 0, we performed a set of
trials for different values of $N$, starting from $N$ initial
conditions (for a system of size $N$) and averaging the data on
$nreal=1000$ disorder realizations. We measured two quantities to
test the performance of the algorithms:
\begin{itemize}
\item [-] the average time (i.e. the number of spin flips) to
reach a minimum energy level
\begin{equation}
\tau=\frac 1M\sum_{i=1}^{M}t_i,
\end{equation}
with $M=N\cdot nreal$ and $t_i$, $i=1,\ldots,M$ the time for each
initial condition;
\item [-] the lowest energy found (averaged over disorder)
\begin{equation}
H_N=\left\langle\frac{\min_{\sigma}H_N(J,\sigma)}{N}\right\rangle_{nreal},
\end{equation}
where $\min_{\sigma}H_N(J,\sigma)$ is the minimum value of the
energy of the me\-ta\-sta\-ble states attained starting from the
set of the $N$ initial conditions.
\end{itemize}
Our numerical experiments follows two different protocols:
\begin{enumerate}
\item with a fixed number of initial conditions;
\item with a fixed elapsed computer time.
\end{enumerate}
The results are described in the following subsections.
\begin{figure}
    \setlength{\unitlength}{1cm}
          \centering
               \includegraphics[width=13cm,height=10cm]{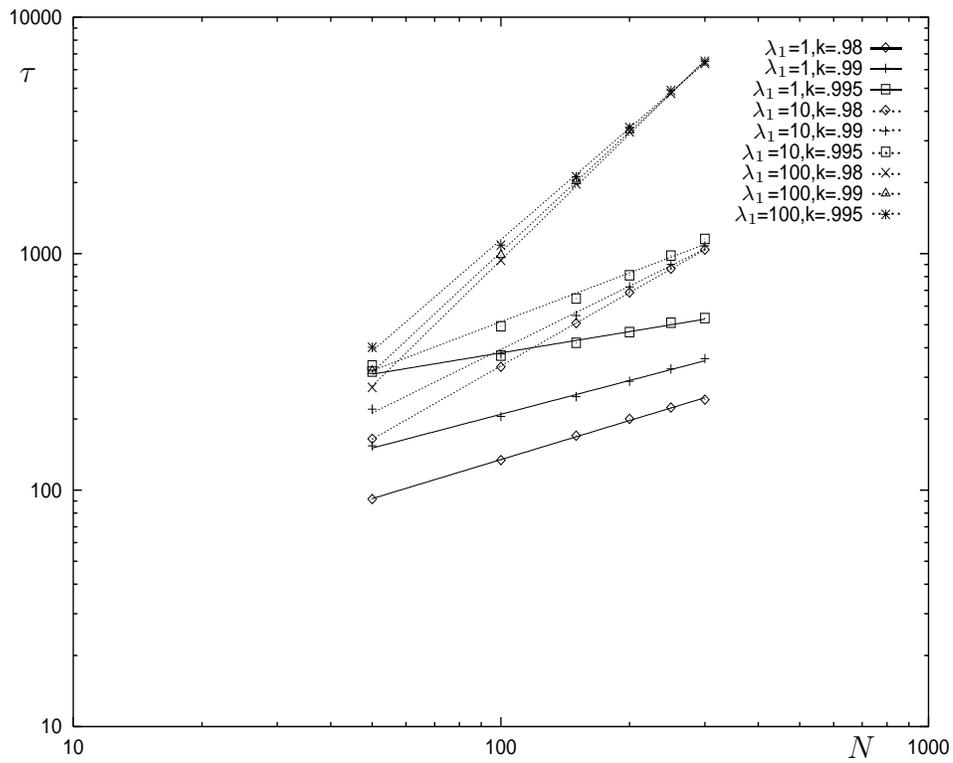}
               \put(-1.5,0){$N$}
               \put(-12.5,9){$\tau$}
               \put(-2.6,9.36){{\scriptsize{$\lambda_1$}}}
               \put(-2.6,9.1){{\scriptsize{$\lambda_1$}}}
               \put(-2.76,8.81){{\scriptsize{$\lambda_1$}}}
               \put(-2.76,8.53){{\scriptsize{$\lambda_1$}}}
               \put(-2.75,8.24){{\scriptsize{$\lambda_1$}}}
               \put(-2.9,7.97){{\scriptsize{$\lambda_1$}}}
               \put(-2.9,7.70){{\scriptsize{$\lambda_1$}}}
               \put(-2.9,7.42){{\scriptsize{$\lambda_1$}}}
               \put(-3,7.14){{\scriptsize{$\lambda_1$}}}
               \thinlines
               \put(-1.18,9.44){\line(1,0){.4}}
               \put(-1.18,9.17){\line(1,0){.4}}
               \put(-1.18,8.89){\line(1,0){.4}}
               \put(-1.18,8.54){\scriptsize{$\cdot$$\cdot$$\cdot$$\cdot$$\cdot$}}
               \put(-1.18,8.28){\scriptsize{$\cdot$$\cdot$$\cdot$$\cdot$$\cdot$}}
               \put(-1.19,7.98){\scriptsize{$\cdot$$\cdot$$\cdot$$\cdot$$\cdot$}}
               \put(-1.19,7.7){\scriptsize{$\cdot$$\cdot$$\cdot$$\cdot$$\cdot$}}
               \put(-1.19,7.44){\scriptsize{$\cdot$$\cdot$$\cdot$$\cdot$$\cdot$}}
               \put(-1.19,7.17){\scriptsize{$\cdot$$\cdot$$\cdot$$\cdot$$\cdot$}}
               \caption{Average time $\tau$ to reach a metastable
               configuration as a function of $N$ for different
               values of $\lambda_1$ and $k$ for Algorithm 2 and for
               a fixed number of initial spin configurations.}\label{fi:CR1new}
\end{figure}

\subsection{Fixed number of initial conditions}
The dynamics of Algorithm 0 has been shown \cite{CGGUV1} to behave as
a smooth interpolation between greedy and reluctant dynamics \cite{BCDG}
depending on the parameter $\lambda$: small $\lambda$ (say $\lambda\simeq 1$)
plays the role of the greedy algorithm, while large $\lambda$ (say
$\lambda\simeq 100$) that of reluctant.
In fact, the relaxation time $\tau(N)$ grows linearly with the system size
when $\lambda\simeq 1$ and quadratically when $\lambda\simeq 100$
(see Tab.~\ref{TabfitCRnew}), as it
was previously observed in \cite{BCDG} for deterministic greedy and
reluctant regimes.

In Fig.~\ref{fi:CR1new}, which refers to Algorithm 2, we represent
$\tau$ as a function of $N$ ($N\in[25,300]$). We performed the analysis
for different values of the control parameters. For the sake of
space, we show only the values
$\lambda_1=1,10,100$ and three values of $k$
($k=.98,.99,.995$) for each $\lambda_1$, together with the best
numerical fits.
Fig.~\ref{fi:CR1new} shows the progressive increase of the slope
in log-log scale from a sub-linear law in $N$ for $\lambda_1=1$
and $k=.98$ ( $\diamond$ \hskip -.51cm ---) to a super-linear one
for $\lambda_1=100$ and $k=.98$ ( $\times$ \hskip -.65cm $\cdots$).
More in detail, the numerical fits of $\tau_{\lambda_1,k}(N)\sim
N^{a}$ in Fig.~\ref{fi:CR1new} are reported in
Tab.\ref{TabfitCRnew}.

\begin{table}[tbh]
\caption{Numerical fits of $\tau_{\lambda}(N)\sim N^{a}$ for Algorithm 0
(with the symbols of Fig.~\ref{fi:SR_CR1}) and of $\tau_{\lambda_1,k}(N) \sim N^{a}$ for Algorithm 1
and  Algorithm 2 (with the symbols of Fig.~\ref{fi:CR1new})} \label{TabfitCRnew}
\begin{center}
\begin{tabular}{|c|c|c||c|c|c||c|c|c|c|}\hline
\multicolumn{3}{|c||}{Alg 0} &\multicolumn{3}{|c||} {Alg 1} & \multicolumn{4}{|c|}{Alg 2} \\ \hline
$\lambda$ & $a$ & symbol & $\lambda_1$ & $k$ & $a$  &$\lambda_1$ & $k$& $a$ & symbol \\ \hline
    &       &     &      & .98  &  .687 &      & .98  &  .549 & $\diamond$ \hskip -.49cm --- \\ \cline{5-6} \cline{8-10}
 1  & 1.027 & $\ast$ &  1   & .99  &  .630 &  1   & .99  &  .475 & $+$ \hskip -.51cm --- \\ \cline{5-6}\cline{8-10}
    &       &     &      & .995 &  .592 &      & .995 &  .299 & $\square$ \hskip -.51cm --- \\ \hline
    &       &     &      & .98  & 1.041 &      & .98  & 1.030 & $\diamond$ \hskip -.49cm $\cdots$ \\ \cline{5-6}\cline{8-10}
10  & 1.263 &     &  10  & .99  &  .948 &  10  & .99  &  .891 & $+$ \hskip -.54cm $\cdots$ \\ \cline{5-6}\cline{8-10}
    &       &     &      & .995 &  .858 &      & .995 &  .687 & $\square$ \hskip -.54cm $\cdots$ \\ \hline
    &       &     &      & .98  & 1.724 &      & .98  & 1.771 & $\times$ \hskip -.54cm $\cdots$ \\ \cline{5-6}\cline{8-10}
100 & 1.932 & $\diamond$ &  100 & .99  & 1.591 &  100 & .99  & 1.691 & $\triangle$ \hskip -.56cm $\cdots$ \\ \cline{5-6}\cline{8-10}
    &       &     &      & .995 & 1.499 &      & .995 & 1.567 & $\ast$ \hskip -.49cm $\cdots$ \\ \hline
\end{tabular}
\end{center}
\end{table}

\begin{figure}
    \setlength{\unitlength}{1cm}
          \centering
               \includegraphics[width=13cm,height=10cm]{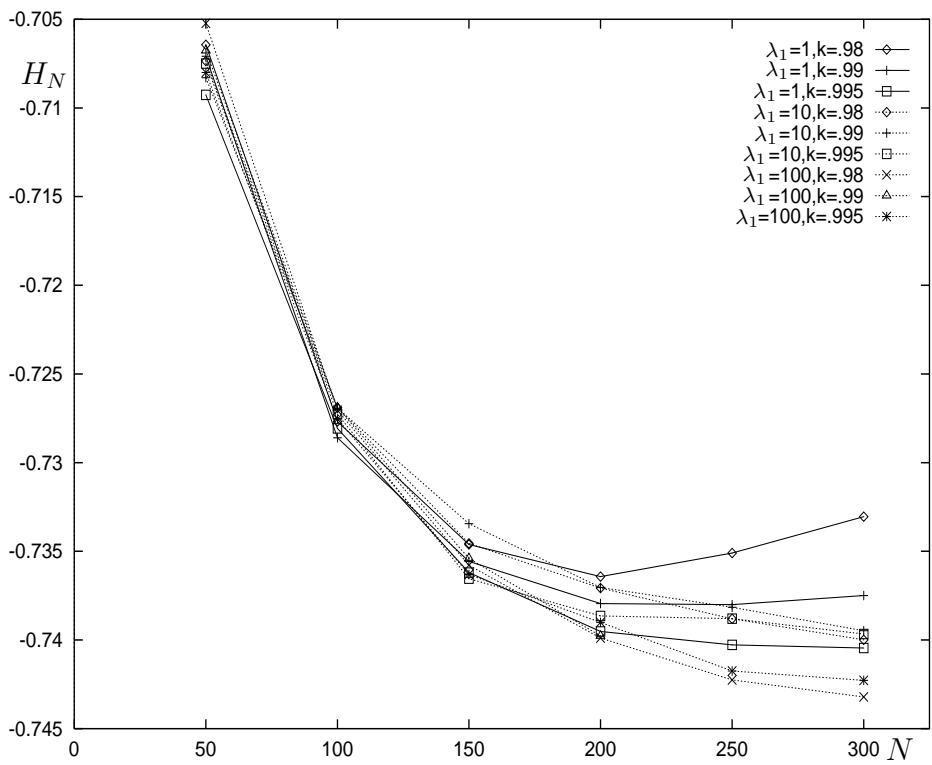}
               \put(-1.,0){$N$}
               \put(-12.5,9){$H_N$}
               \put(-2.6,9.36){{\scriptsize{$\lambda_1$}}}
               \put(-2.6,9.1){{\scriptsize{$\lambda_1$}}}
               \put(-2.76,8.81){{\scriptsize{$\lambda_1$}}}
               \put(-2.76,8.53){{\scriptsize{$\lambda_1$}}}
               \put(-2.75,8.24){{\scriptsize{$\lambda_1$}}}
               \put(-2.9,7.97){{\scriptsize{$\lambda_1$}}}
               \put(-2.9,7.70){{\scriptsize{$\lambda_1$}}}
               \put(-2.9,7.42){{\scriptsize{$\lambda_1$}}}
               \put(-3.,7.14){{\scriptsize{$\lambda_1$}}}
               \caption{Lowest energy value $H_N$ as a function of $N$ for different
               values of $\lambda_1$ and $k$ for Algorithm 2 and for a fixed number
               of initial conditions.}\label{fi:CR2new}
\end{figure}

\begin{figure}
    \setlength{\unitlength}{1cm}
         \centering
               \includegraphics[width=13cm,height=10cm]{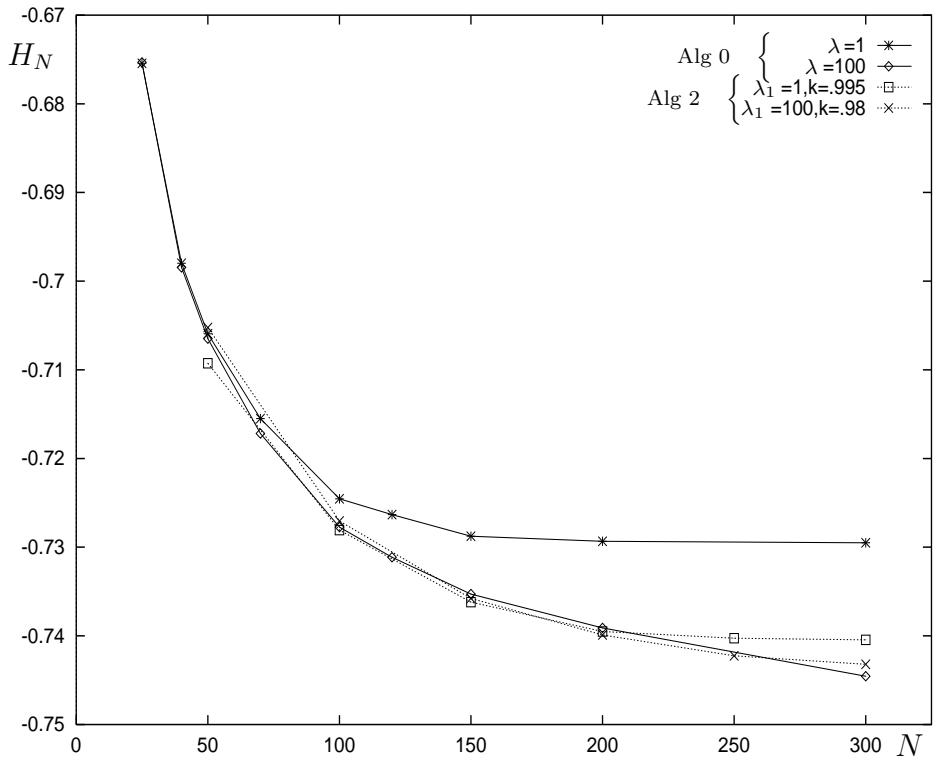}
               \put(-0.9,0){$N$}
               \put(-12.7,9.2){$H_N$}
               \put(-1.8,9.36){{\scriptsize{$\lambda$}}}
               \put(-2.1,9.08){{\scriptsize{$\lambda$}}}
               \put(-2.8,8.8){{\scriptsize{$\lambda_1$}}}
               \put(-2.95,8.52){{\scriptsize{$\lambda_1$}}}
               \put(-3.8,9.22){{\scriptsize{$\textrm{Alg 0}\quad \left\{ \begin{array}{ll}
               & \\
               &
               \end{array}\right.$}}}
               \put(-4.2,8.66){{\scriptsize{$\textrm{Alg 2}\quad \left\{ \begin{array}{ll}
               & \\
               &
               \end{array}\right.$}}}
               \caption{Lowest energy value $H_N$ as a function of $N$ obtained using
               a protocol with a fixed number of initial conditions for $\lambda=1$ ($\ast$)
               and $\lambda=100$ ($\diamond$) for Algorithm 0 and for $\lambda_1=1$ and $k=.995$ ($\square$)
               and for $\lambda_1=100$ and $k=.98$ ($\times$) for Algorithm 2.}
                \label{fi:SR_CR2}
\end{figure}

With the same protocol (fixed number of initial conditions), we
measured the lowest energy $H_N$ found by the algorithms.
As a general remark we recall that from a theoretical
point of view it is proved the monotonicity in $N$ of the
ground state energy (this follows from sub-additivity \cite{GT}).
For the largest size we have studied, some values of the
simulation parameters give a non-monotone behavior in
$N$, suggesting that we are not actually finding the
true lowest energy state. A larger number of trials
(i.e. initial conditions) would be needed to achieve
the global minimum. However, our principal aim here
is not to have a perfect measure of ground state energies.
In Fig.~\ref{fi:CR2new} we represent, for Algorithm 2, $H_N$ as a
function of $N$ for different values of $\lambda_1$ and $k$. The
best results for large $N$ are obtained for $\lambda_1=100$ and
$k=.98$ which corresponds to annealed reluctant dynamics (as
found for Algorithm 0, see Fig.~\ref{fi:SR_CR2}).
Therefore, this confirms \cite{BCGGUV,CGGUV1} that, for a fixed
number of initial spin configurations, the algorithm that makes
moves corresponding to the ``smallest'' possible energy change
keeping the possibility of energy increase only for the first
steps of the algorithm is the most efficient in reaching
low-energy states. Note that, for $\lambda_1=1$ and $k=.995$ the
attained energy values are sufficiently low: even if these results
are not better than those for $\lambda_1=100$ (with $k=.98$ and $k=.995$),
they should not be discarded since the average time scales better (
$\tau_{1,.995}^{(2)}(N)\sim N^{.299}$ instead of $\tau_{100,.98}^{(2)}(N)\sim N^{1.771}$
or $\tau_{100,.995}^{(2)}(N)\sim N^{1.567}$)
\footnote{From now on, the superscript $(x)$ in the notation of the average time $\tau^{(x)}$ will
refer to the number of the corresponding algorithm.}.

\begin{figure}
    \setlength{\unitlength}{1cm}
          \centering
               \includegraphics[width=13cm,height=10cm]{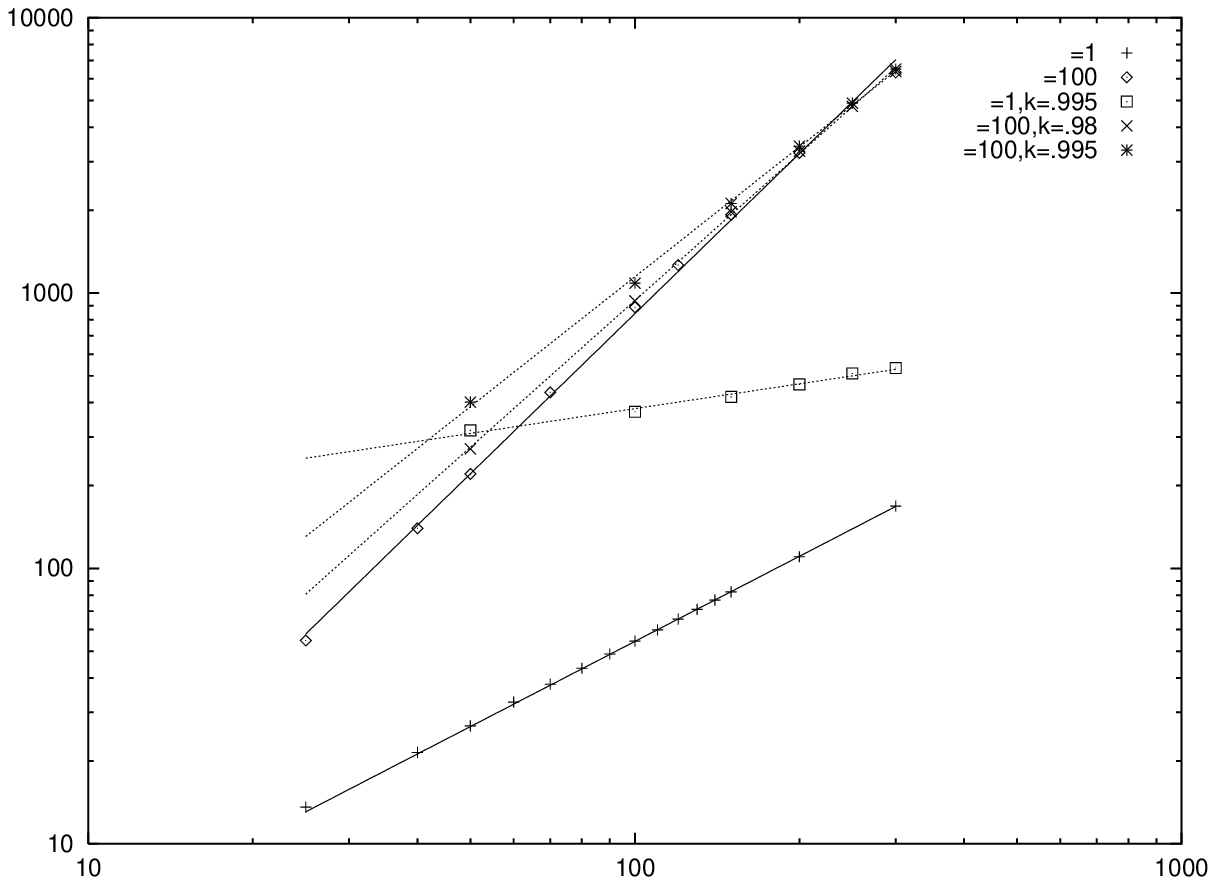}
               \put(-1.5,0){$N$}
               \put(-12.5,9){$\tau$}
               \put(-1.8,9.36){{\scriptsize{$\lambda$}}}
               \put(-2.1,9.1){{\scriptsize{$\lambda$}}}
               \put(-2.8,8.79){{\scriptsize{$\lambda_1$}}}
               \put(-2.9,8.53){{\scriptsize{$\lambda_1$}}}
               \put(-3.1,8.25){{\scriptsize{$\lambda_1$}}}
               \put(-11,4){{\scriptsize{$\textrm{Alg 2}\quad \left\{ \begin{array}{ll}
               & \\
               & \\
               & \\
               & \\
               & \\
               &
               \end{array}\right.$}}}
               \put(-11,2.7){{\scriptsize{$\textrm{Alg 0}\quad \left\{ \right.$}}}
               \put(-11,.7){{\scriptsize{$\textrm{Alg 0}\quad \left\{ \right.$}}}
               \caption{Average time $\tau$ to reach a metastable
               configuration as a function of $N$ for $\lambda=1$ ($+$) and for
               $\lambda=100$ ($\diamond$) for Algorithm 0, and for $\lambda_1=1$ and $k=.995$ ($\square$),
               for $\lambda_1=100$ and $k=.98$ ($\times$),and for $\lambda_1=100$ and $k=.995$ (
               $\ast$) for Algorithm 2.}
               \label{fi:SR_CR1}
\end{figure}

Comparing these results with those obtained with the interpolating greedy and reluctant algorithm
(Algorithm 0) \cite{CGGUV1} we note (Figs.~\ref{fi:SR_CR2} and ~\ref{fi:SR_CR1} and Tab.~\ref{TabfitCRnew})
that for small $\lambda$ and $\lambda_1$ Algorithm 2 is better performing
than Algorithm 0 both with respect to average time and energy levels, while
for greater $\lambda$ and $\lambda_1$ we find comparable energy
values but with lower cost for the computational time for Algorithm 2
($\tau_{100,.98}^{(2)}(N)\sim N^{1.771}$ instead of $\tau_{100}^{(0)}(N)\sim
N^{1.932}$).

\begin{figure}
    \setlength{\unitlength}{1cm}
          \centering
               \includegraphics[width=13cm,height=10cm]{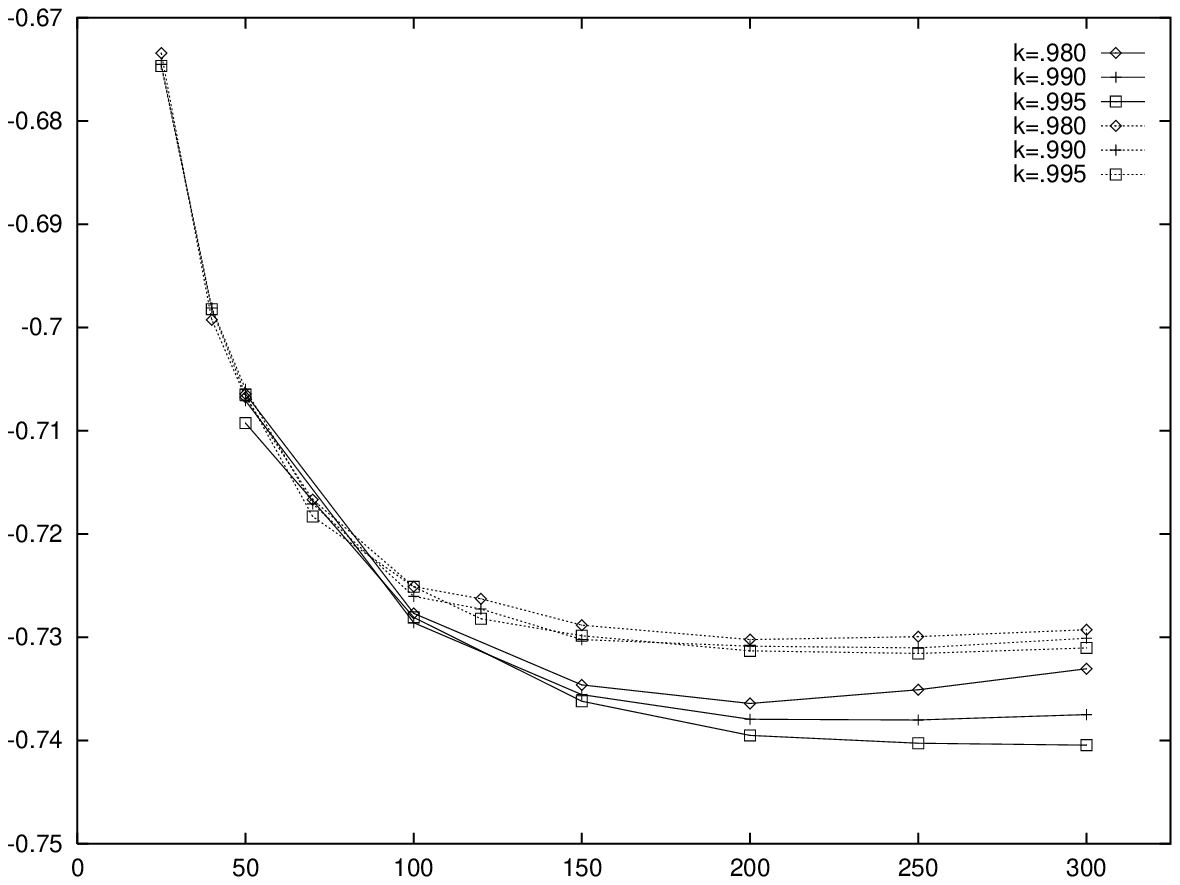}
               \put(-1,0){$N$}
               \put(-12.5,9){$H_N$}
               \put(-4.3,9.1){{\scriptsize{$\textrm{Alg 2,}\quad \lambda_1=1\left\{ \begin{array}{ll}
               & \\
               & .
               \end{array}\right.$}}}
               \put(-4.3,8.24){{\scriptsize{$\textrm{Alg 1,}\quad \lambda_1=1\left\{ \begin{array}{ll}
               & \\
               & .
               \end{array}\right.$}}}
               \caption{Lowest energy value $H_N$ as a function of $N$ for
               $\lambda_1=1$ and different values of $k$, for Algorithm 1 and 2.}\label{fi:CR1}
\end{figure}

\begin{figure}
    \setlength{\unitlength}{1cm}
          \centering
               \includegraphics[width=13cm,height=10cm]{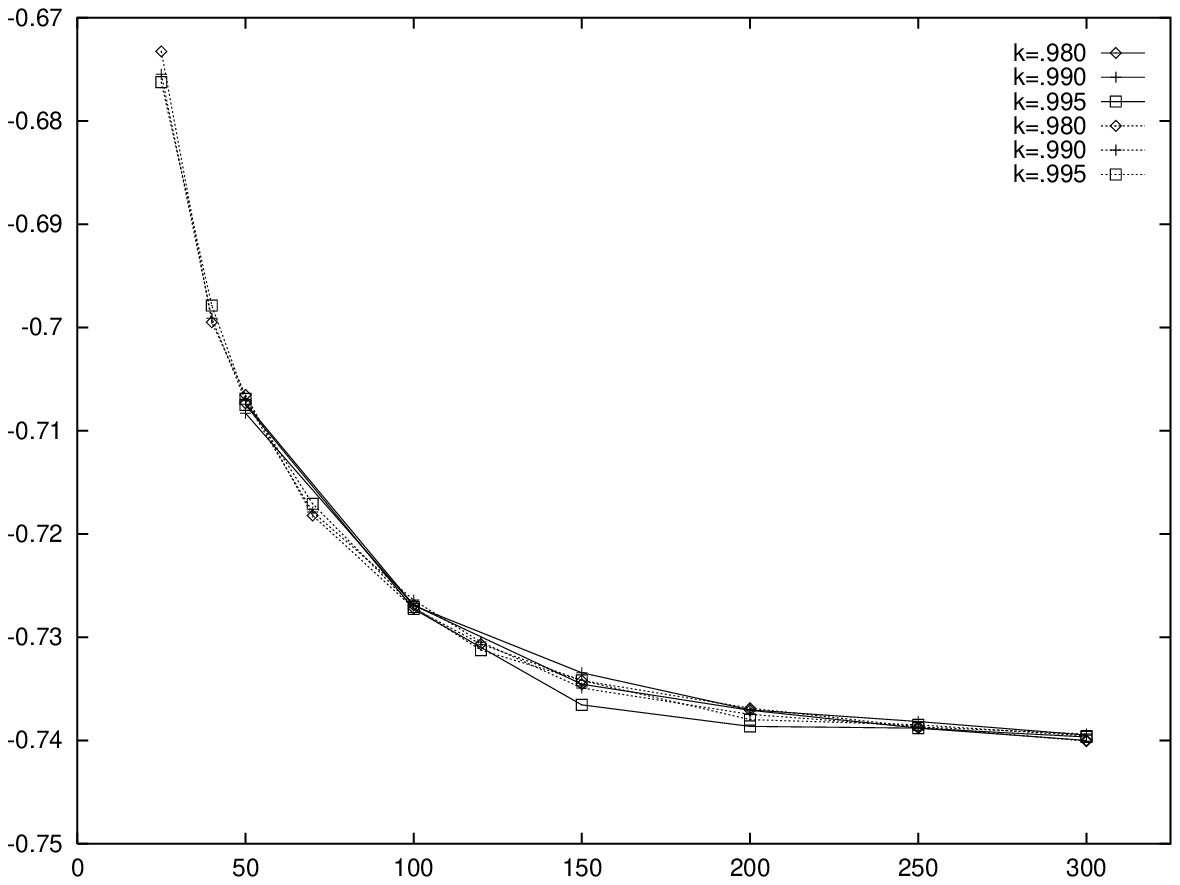}
               \put(-1.,0){$N$}
               \put(-12.5,9){$H_N$}
               \put(-4.5,9.1){{\scriptsize{$\textrm{Alg 2,}\quad \lambda_1=10\left\{ \begin{array}{ll}
               & \\
               & .
               \end{array}\right.$}}}
               \put(-4.5,8.24){{\scriptsize{$\textrm{Alg 1,}\quad \lambda_1=10\left\{ \begin{array}{ll}
               & \\
               & .
               \end{array}\right.$}}}
               \caption{Lowest energy value $H_N$ as a function of $N$ for
               $\lambda_1=10$ and for different values of $k$ obtained with
               Algorithm 1 and 2.}\label{fi:CR2}
\end{figure}

The same analysis is considered also for Algorithm 1.
The comparison between Algorithms 1 and 2 shows that the possibility of exceed the energy
barriers between minima is useful only for small values of $\lambda_1$ (for
$\lambda_1$ close to $1$ Algorithm $2$ is more efficient than Algorithm $1$ in
reaching lower energy states)
while for $\lambda_1\ge 5$ the performances of Algorithms $1$ and $2$
are practically indistinguishable (see Figs.~\ref{fi:CR1} and \ref{fi:CR2}).
Moreover, we note
that the best scaling of the average time $\tau_{\lambda_1,k}$ with respect to $N$
is obtained with Algorithm $2$ (see Tab.~\ref{TabfitCRnew}),
though for fixed $N,\lambda_1$ and $k$, we have $\tau^{(1)}_{\lambda_1,k}<
\tau^{(2)}_{\lambda_1,k}$.

\begin{figure}
    \setlength{\unitlength}{1cm}
          \centering
               \includegraphics[width=13cm,height=10cm]{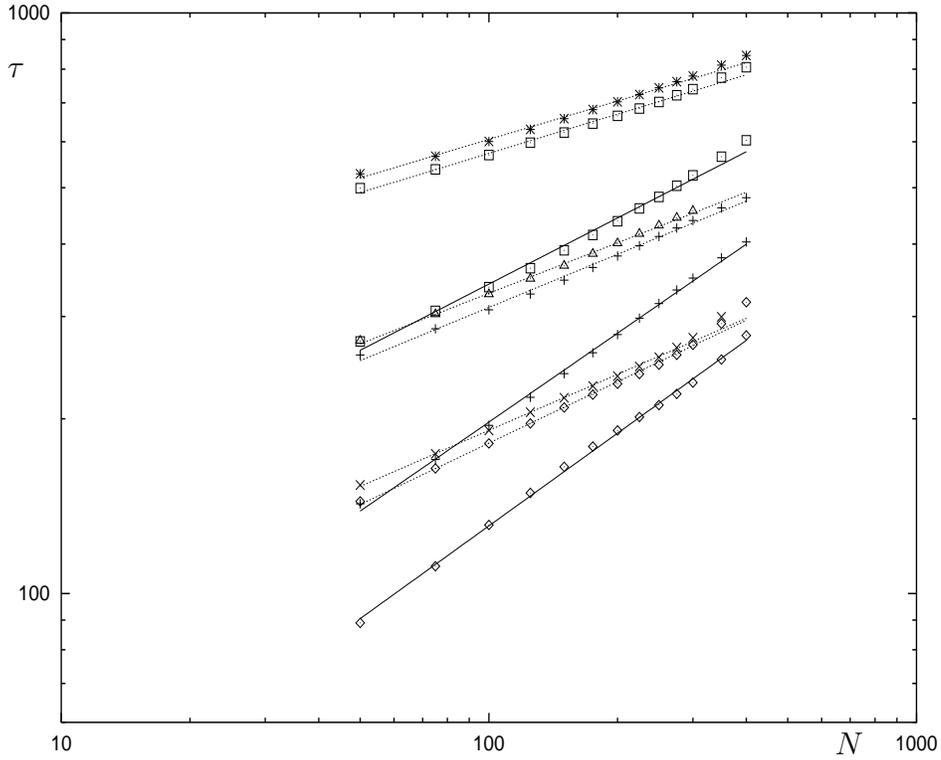}
               \put(-1.5,0){$N$}
               \put(-12.5,9){$\tau$}
               \caption{Average time $\tau$ to reach a metastable
               configuration as a function of $N$ for different
               values of $\lambda_1(0)$ and $k$ for Algorithm 3,
               together with the best numerical fits for a fixed number of initial
               conditions. We represent $\lambda_1(0)=2$ ($k=.98$ ($\diamond$ \hskip -.4cm ---),
               $k=.99$ ($+$ \hskip -.49cm ---) and $k=.995$ ($\square$ \hskip -.49cm ---)),
               $\lambda_1(0)=10$ ($k=.98$ ($\diamond$ \hskip -.45cm $\cdots$),
               $k=.99$ ($+$ \hskip -.49cm $\cdots$) and $k=.995$
               ($\square$ \hskip -.49cm $\cdots$)) and $\lambda_1(0)=100$
               ($k=.98$ ($\times$ \hskip -.49cm $\cdots$), $k=.99$ ($\triangle$
               \hskip -.58cm $\cdots$) and $k=.995$ ($\ast$ \hskip -.49cm $\cdots$))
               }\label{fi:SA1}
\end{figure}

\begin{figure}
    \setlength{\unitlength}{1cm}
          \centering
               \includegraphics[width=13cm,height=10cm]{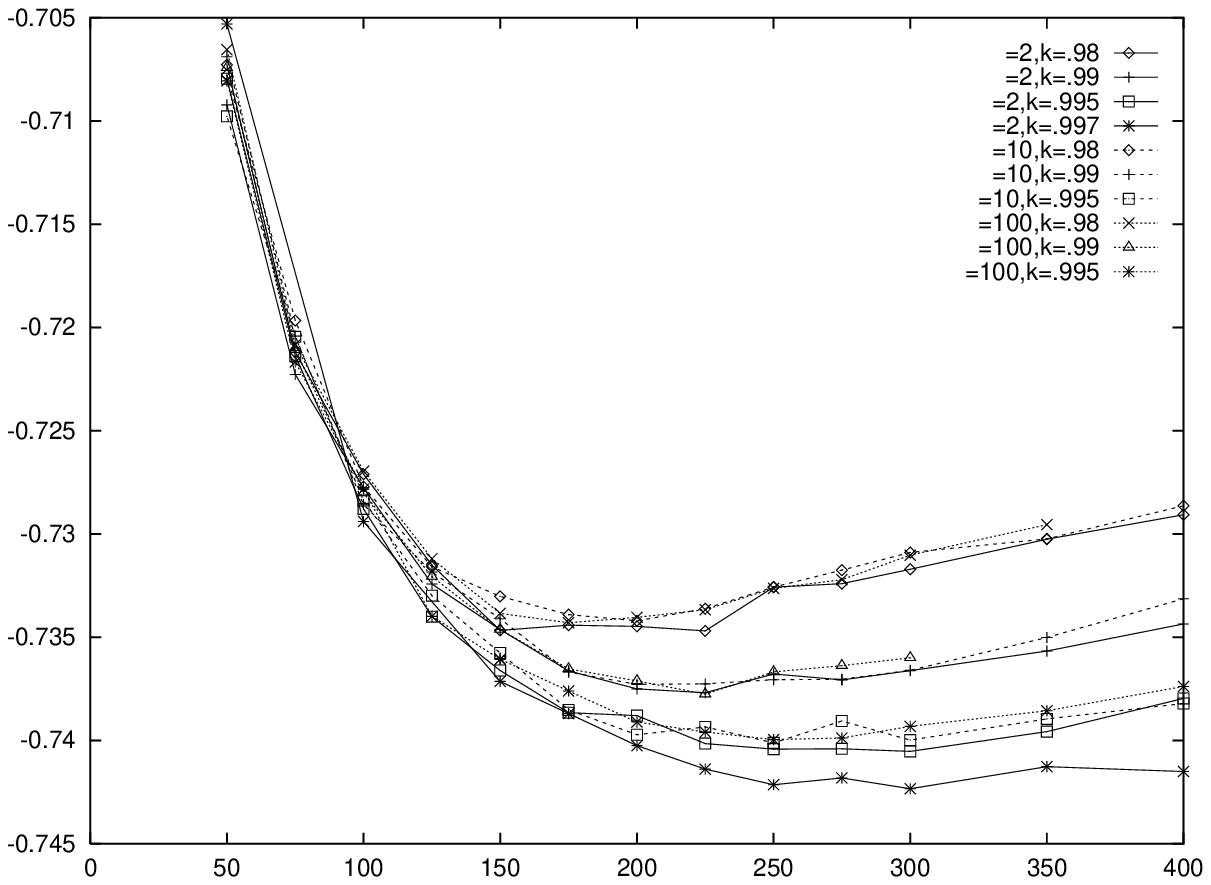}
               \put(-1.,0){$N$}
               \put(-12.5,9){$H_N$}
               \put(-3,9.36){{\scriptsize{$\lambda_1(0)$}}}
               \put(-3,9.1){{\scriptsize{$\lambda_1(0)$}}}
               \put(-3.16,8.81){{\scriptsize{$\lambda_1(0)$}}}
               \put(-3.16,8.53){{\scriptsize{$\lambda_1(0)$}}}
               \put(-3.16,8.24){{\scriptsize{$\lambda_1(0)$}}}
               \put(-3.15,7.97){{\scriptsize{$\lambda_1(0)$}}}
               \put(-3.3,7.70){{\scriptsize{$\lambda_1(0)$}}}
               \put(-3.3,7.42){{\scriptsize{$\lambda_1(0)$}}}
               \put(-3.3,7.14){{\scriptsize{$\lambda_1(0)$}}}
               \put(-3.4,6.86){{\scriptsize{$\lambda_1(0)$}}}
               \caption{Lowest energy value $H_N$ as a function of $N$ for different
               values of $\lambda_1(0)$ and $k$ for Algorithm 3 and for
               a fixed number of initial conditions.}\label{fi:SA2}
\end{figure}

\vskip .2cm\noindent
Figures \ref{fi:SA1} and \ref{fi:SA2} report the results of the analysis
of Algorithm 3 with a fixed number of initial conditions:
$N\in[25,400]$)
for three distinct values of $\lambda_1(0)$
($\lambda_1(0)=2,10,100$) and for four values of $k$
($k=.98,.99,.995,.997$) for each $\lambda_1(0)$. Because of high computational costs (which
increase with $\lambda_1(0)$ and $k$), the cases $N=350$ and $N=400$
for $\lambda_1(0)=100$ are only partially studied. For the same reason also the case
$k=.997$ is considered only for $\lambda_1(0)=2$.
\vskip .2cm\noindent
Fig.~\ref{fi:SA2} shows that Algorithm 3 seems to depend weakly on the
parameter $\lambda_1(0)$, its behavior being mainly ruled by $k$. In fact,
the lines of the $H_N$ values corresponding to the same choices of $k$ are
grouped into narrow bands well separated one from the others. Moreover,
a closer look to Fig.~\ref{fi:SA2} shows that the best result for $H_N$
is obtained for $\lambda_1(0)=2$ and $k=.997$. Note that for any $\lambda_1(0)$,
the closer the values of $k$ to one, the lower the values of energy:
slow growths of the
parameter $\lambda_2(t)$ enable energy increases and then the
possibility to exceed the energy barriers.
Even though Algorithm 2 is slightly better performing ($\lambda_1=100, k=.98$ see
Fig.~\ref{fi:CR2new}) in terms
of minimum energy level reached, the best scaling of $\tau_{\lambda_1(0),k}(N)$ is
obtained by Algorithm 3. In fact, for Algorithm 3 we note (Fig.~\ref{fi:SA1}
and Tab.~\ref{TabfitSA}) the progressive increase of the slope
in log-log scale from a scaling law $\tau_{\lambda_1(0),k}^{(3)}(N)\sim
N^{.22}$ for $\lambda_1(0)=100$
and $k=.995$ ( $\ast$ \hskip -.51cm $\cdots$) to $\tau_{\lambda_1(0),k}^{(3)}(N)\sim
N^{.53}$
for $\lambda_1(0)=2$ and $k=.98$ ( $\diamond$ \hskip -.49cm ---).
More in detail, the numerical fits of $\tau_{\lambda_1(0),k}(N)\sim
N^{a}$ for Algorithm 3 are reported in Tab.\ref{TabfitSA}.

\begin{figure}
    \setlength{\unitlength}{1cm}
          \centering
               \includegraphics[width=13cm,height=10cm]{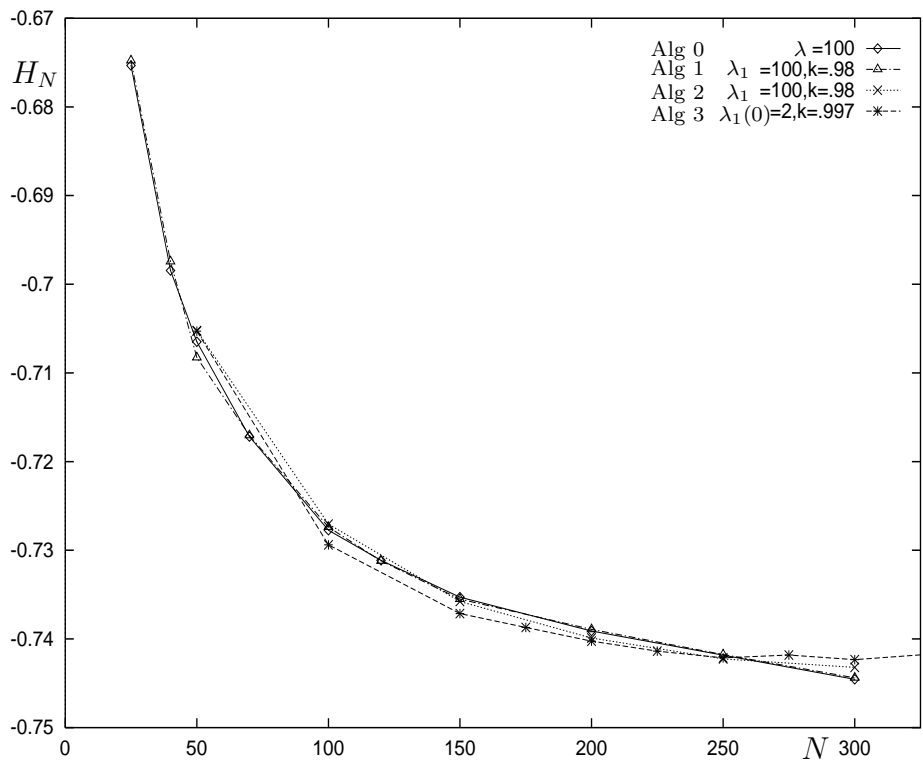}
               \put(-2,0){$N$}
               \put(-12.5,9){$H_N$}
               \put(-2.1,9.36){{\scriptsize{$\lambda$}}}
               \put(-4,9.36){{\scriptsize{Alg 0}}}
               \put(-3,9.1){{\scriptsize{$\lambda_1$}}}
               \put(-4,9.1){{\scriptsize{Alg 1}}}
               \put(-3,8.79){{\scriptsize{$\lambda_1$}}}
               \put(-4,8.79){{\scriptsize{Alg 2}}}
               \put(-3.1,8.49){{\scriptsize{$\lambda_1(0)$}}}
               \put(-4,8.49){{\scriptsize{Alg 3}}}
               \caption{Lowest energy value $H_N$ as a function of $N$ for $\lambda=100$ ($\diamond$) for Algorithm 0,
               for $\lambda_1=100$ and $k=.98$ ($\triangle$) for Algorithm 1 and ($\times$)
               for Algorithm 2 and for $\lambda_1(0)=2$ and $k=.997$ ($\ast$) for Algorithm 3.}
               \label{fi:SR_SA}
\end{figure}

To conclude the analysis of the protocol with a fixed
number of initial conditions we can say that  taking into account
also the average time $\tau$, the best performing algorithm
in reaching minimum energy level is Algorithm 3 (Fig.~\ref{fi:SR_SA}). In fact,
Algorithm 3 with $\lambda_1(0)=2$ e $k=.997$ attains minimum energy levels comparable
with those obtained by the other algorithms with $\lambda$ and $\lambda_1$ equal to $100$
but with lower computational costs ($\tau_{2,.997}^{(3)}\sim N^{.272}$ while
$\tau_{100}^{(0)}\sim N^{1.932}$, $\tau_{100,.98}^{(1)}\sim N^{1.724}$ and
$\tau_{100,.98}^{(2)}\sim N^{1.771}$, see Tabs.~\ref{TabfitCRnew}
and \ref{TabfitSA}).

\subsection{Fixed elapsed computer time}
\begin{figure}
    \setlength{\unitlength}{1cm}
          \centering
               \includegraphics[width=13cm,height=10cm]{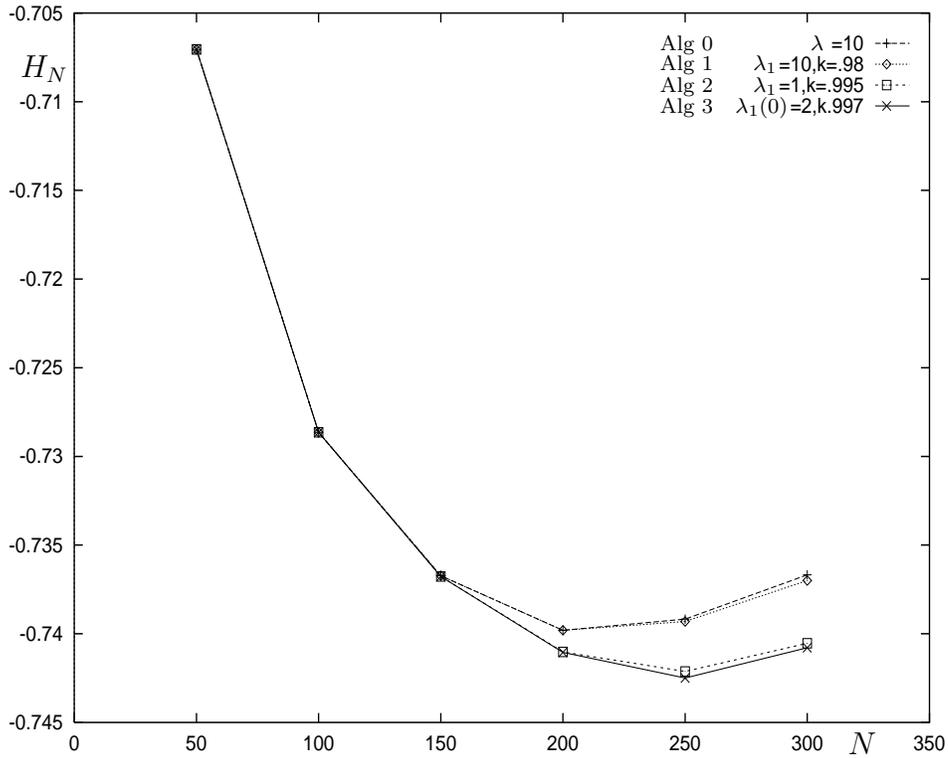}
               \put(-1.5,0){$N$}
               \put(-12.5,9){$H_N$}
               \put(-4,9.36){\scriptsize{Alg 0}}
               \put(-2,9.36){{\scriptsize{$\lambda$}}}
               \put(-4,9.1){\scriptsize{Alg 1}}
               \put(-2.76,9.1){{\scriptsize{$\lambda_1$}}}
               \put(-4,8.81){\scriptsize{Alg 2}}
               \put(-2.76,8.81){{\scriptsize{$\lambda_1$}}}
               \put(-4,8.53){\scriptsize{Alg 3}}
               \put(-3,8.53){{\scriptsize{$\lambda_1(0)$}}}
               \caption{Lowest energy value $H_N$  as a function of $N$ for different
               values of control parameters for Algorithms 0, 1, 2 and 3, for a fixed
               CPU time of $50$ h on a IBM SP4. The symbol (+) refers to
               $\lambda=10$ for Algorithm 0, $(\diamond)$ to $\lambda_1=10$ and $k=.98$ for Algorithm 1, ($\square$)
               to $\lambda_1=1$ and $k=.995$ for Algorithm 2 and ($\times$) for $\lambda_1(0)=2$ and $k=.997$
               for Algorithm 3.}\label{fi:CR3new}
\end{figure}

Finally, we analyze the lowest energy states found by the dynamics
varying the control parameters for a given elapsed running time for all
algorithms. In
Fig.~\ref{fi:CR3new} we consider the minimum energy values $H_N$,
obtained by choosing different system sizes $N$ and, for each
of them, different parameter values ($\lambda=1,10,100$ for Algorithm 0,
$\lambda_1=1, 5, 10, 100$ for Algorithm 1, $\lambda_1=1, 10$ for Algorithm 2
and $\lambda_1(0)=2, 10, 100$ for Algorithm 3)
with different annealing scheduling each ($k=.98$ and
$k=.995$ for Algorithm 1 and 2, $k=.995$ and $k=.997$ for Algorithm 3),
for a fixed time of $50$ h of CPU on a IBM SP4.
For Algorithm 2 we consider in detail mainly the case ($\lambda_1=1$) in which
the dynamics behaves differently from that generated by Algorithm 1.
Each run (i.e. for fixed $N$ and for fixed control parameter) consists of $1000$
disorder realizations, with the same CPU time length ($3$ min.)
assigned to each sample, in order to compare these results with
\cite{BCDG,BCGGUV,CGGUV1}. With all this dynamics, for $N\le 150$, we
believe to find the ground state of the system, since varying
the control parameters and independently on the algorithm used,
the values of $H_N$ coincide, within our
numerical accuracy ($10^{-10}$). The
best result is obtained with Algorithm 3 for the case $\lambda_1(0)=2$
and $k=.997$ (even though the result provided by Algorithm 2 for
$\lambda_1=1$ and $k=.995$ is comparable). Note that, for Algorithm 1 the best
result is for $\lambda_1=10$ and $k=.98$ in good agreement
with the best result of Algorithm 0 obtained for $\lambda=10$ (Fig.~\ref{fi:CR3new}).
Moreover, it is worthnoting that the values $H_N$ obtained
with Algorithm 3  for the case $\lambda_1(0)=2$
and $k=.997$ are the best (for fixed CPU time) with respect to all algorithms
we consider in the present paper and in \cite{BCDG,BCGGUV,CGGUV1}.
\vskip .5cm

\begin{table}[tbh]
\caption{Numerical fits of $\tau_{\lambda_1(0),k}(N) \sim
N^{a}$ for Algorithm 3} \label{TabfitSA}
\begin{center}
\begin{tabular}{|c|c|c|c|}\hline
$\lambda_1(0)$ & $k$ & $a$ & symbol \\
\hline
       & .98  &  .531 & $\diamond$ \hskip -.49cm --- \\ \cline{2-4}
   2   & .99  &  .509 & $+$ \hskip -.51cm --- \\ \cline{2-4}
       & .995 &  .379 & $\square$ \hskip -.51cm --- \\ \cline{2-4}
       & .997 &  .272 & $\ast$ \hskip -.49cm --- \\ \hline
       & .98  &  .352 & $\diamond$ \hskip -.49cm $\cdots$ \\ \cline{2-4}
  10   & .99  &  .304 & $+$ \hskip -.54cm $\cdots$ \\ \cline{2-4}
       & .995 &  .225 & $\square$ \hskip -.54cm $\cdots$ \\ \hline
       & .98  &  .321 & $\times$ \hskip -.54cm $\cdots$ \\ \cline{2-4}
  100  & .99  &  .289 & $\triangle$ \hskip -.56cm $\cdots$ \\ \cline{2-4}
       & .995 &  .220 & $\ast$ \hskip -.49cm $\cdots$ \\ \hline
\end{tabular}
\end{center}
\end{table}

\newpage

\section{Acknowledgments}
We thank Prof. S. Graffi and Prof. I. Galligani for their encouragement.
The Cineca staff and in particular Dr. G. Erbacci and Dr. C. Calonaci are
acknowledged for the technical support.
The computation resources were provided
by Cineca (High Performance Computing Grant) and by
CICAIA (Universit\`a di Modena e Reggio Emilia).


\begin{thebibliography}{99}

\bibitem{SK}
D. Sherrington S. Kirkpatrick,
``Solvable Model of a Spin-Glass''
{\em Phys. Rev. Lett.} {\bf 35} 1792-1796 (1975).

\bibitem{MPV} M. M\'ezard, G. Parisi, and M. A. Virasoro,
{\it Spin Glass Theory and Beyond,}
(World  Scientific, Singapore, 1987).

\bibitem{Bouch} J.-P. Bouchaud , M. Potters,
{\em Theory of Financial Risk}, Alea-Saclay, Eyrolles, Paris (1997).

\bibitem{Ni} H. Nishimori,
{\em Statistical Physics of Spin Glasses and
Information Processing}, Oxford University Press, New York (2001).

\bibitem{BP}
F. T. Bantilan and R. G. Palmer,
``Magnetic properties of a model spin glass and the failure
of linear response theory'',
{\em J. Phys. F} {\bf 11} 261-266 (1981).

\bibitem{CMPP}
S. Cabasino, E. Marinari, P. Paolucci and G. Parisi,
``Eigenstates and limit cycles in the SK model''
{\em J. Phys. A: Math. Gen.} {\bf 21} 4201-4210 (1988).

\bibitem{KGV}
S. Kirkpatrick, C.D. Gelatt, M.P. Vecchi,
{\em Science} {\bf 220} 671 (1983).

\bibitem{GSL}
G.S. Grest, C.M. Soukoulis, K. Levin,
``Cooling-rate dependence for the spin-glass ground-state energy:
implications for optimization by simulated annealing'',
{\em Pys. Rev. Lett.} {\bf 56} 1148-1151 (1986).

\bibitem{BKM}
 J.-P. Bouchaud, F. Krzakala, and O. C. Martin,
``Energy exponents and corrections to scaling in Ising spin glasses'',
{\em Phys. Rev. B} {\bf 68}, 224404 (2003).

\bibitem{P}
M. Palassini,
``Ground-state energy fluctuations in the Sherrington-Kirkpatrick model'',
cond-mat/0307713.

\bibitem{BP01}
S. Boettcher, A.G. Percus,
``Optimization with Extremal Dynamics'',
{\em Phys. Rev. Lett.} {\bf  86}  5211-5214 (2001).

\bibitem{BS}
S. Boettcher, P. Sibani
``Comparing extremal and thermal explorations of energy
landscapes'',
cond-mat/0406543.

\bibitem{B}
S. Boettcher,
``Extremal Optimization for the Sherrington-Kirkpatrick Spin Glass'',
cond-mat/0407130.

\bibitem{BCDG}
L.Bussolari, P. Contucci, M. Degli Esposti, C. Giardin\`a
``Energy-Decreasing Dynamics in Mean-Field Spin Models'' {\em
Jour. Phys. A: Math. Gen.}  {\bf 36}  2413-2421 (2003).

\bibitem{BCGGUV}
L. Bussolari, P.Contucci, C. Giardin\`a, C. Giberti, F.
Unguendoli, C. Vernia, ``Optimization strategies in complex
systems'', {\em Science and Supercomputing at Cineca -  2003
Report}, 386-390, http://arxiv.org/abs/math.NA/0309058.

\bibitem{CGGUV1}
P.Contucci, C. Giardin\`a, C. Giberti, F. Unguendoli, C. Vernia,
``Interpolating greedy and reluctant algorithms'', to appear on
{\it Optimization Methods and Software} (2004),
http://arxiv.org/abs/math-ph/0309063.

\bibitem{GT} F. Guerra and F. Toninelli,
``The thermodynamical limit in mean field
spin glass model'',
{\it Commun. Math. Phys.} {\bf 230}, 71-79, (2002).





\end{thebibliography}
\end{document}